\documentclass{article}

\usepackage{PRIMEarxiv}

\usepackage[utf8]{inputenc} 
\usepackage[T1]{fontenc}    
\usepackage{hyperref}       
\usepackage{url}            
\usepackage{booktabs}       
\usepackage{amsfonts}       
\usepackage{nicefrac}       
\usepackage{microtype}      
\usepackage{lipsum}
\usepackage{fancyhdr}       
\usepackage{graphicx}       
\usepackage{amsmath}
\usepackage{xcolor}
\usepackage{soul}
\usepackage{hyperref}

\usepackage{mathtools}
\usepackage{listings}
\usepackage{cite}
\usepackage{amsmath,amssymb,amsfonts}
\usepackage{algorithmic}
\usepackage{graphicx}
\usepackage{textcomp}
\usepackage{xcolor}
\def\BibTeX{{\rm B\kern-.05em{\sc i\kern-.025em b}\kern-.08em
    T\kern-.1667em\lower.7ex\hbox{E}\kern-.125emX}}

\makeatletter
\newcommand*{\rom}[1]{\expandafter\@slowromancap\romannumeral #1@}
\makeatother
\usepackage{titlesec}
\titleclass{\subsubsubsection}{straight}[\subsection]
\titleformat{\subsubsubsection}
  {\normalfont\normalsize\bfseries}{\thesubsubsubsection}{1em}{}
\titlespacing*{\subsubsubsection}
{0pt}{3.25ex plus 1ex minus .2ex}{1.5ex plus .2ex}
\graphicspath{{media/}}     

\title{LAKEE: A Lightweight Authenticated Key Exchange Protocol for Power Constrained Devices
}

\author{
  Seyedsina Nabavirazavi \\
  Florida International University \\
  \texttt{\{snabavir@fiu.edu} \\
   \And
  S. Sitharama Iyengar \\
  Florida International University \\
  \texttt{iyengar@cs.fiu.edu} \\
}

\begin{document}
\maketitle

\begin{abstract}
The rapid development of IoT networks has led to a research trend in designing effective security features for them. Due to the power-constrained nature of IoT devices, the security features should remain as lightweight as possible. Currently, most of the IoT network traffic is unencrypted. The leakage of smart devices' unencrypted data can come with the significant cost of a privacy breach. To have a secure channel with encrypted traffic, two endpoints in a network have to authenticate each other and calculate a short-term key. They can then communicate through an authenticated and secure channel. This process is referred to as authenticated key exchange (AKE). Although Datagram Transport Layer Security (DTLS) offers an AKE protocol for IoT networks, research has proposed more efficient and case-specific alternatives. This paper presents LAKEE, a straightforward, lightweight AKE protocol for IoT networks. Our protocol employs elliptic curve cryptography for generating a short-term session key. It reduces the communication and computational overhead of its alternatives while maintaining or improving their security strength. The simplicity and low overhead of our protocol make it a fit for a network of constrained devices.
\end{abstract}

\keywords{Authenticated Key Exchange \and Authentication \and Session Establishment \and Elliptic Curve Cryptography \and Internet of Things \and Constrained Application Protocol \and IoT \and Protocol Design \and Smart City}

\section{Introduction}
\subsection{The Rise of IoT}
 Security issues of IoT networks have been a focal point for researchers since the rise of IoT. Research has offered various mechanisms for achieving security goals, i.e., confidentiality, integrity, and availability in this field. However, providing security would not be straightforward in this domain. IoT devices are power-constrained and have specific functionality. Using general-purpose protocols for every IoT device would result in more battery consumption and less quality of service. Therefore, numerous protocols have been proposed for each layer of the IoT protocol stack. IoT designers have to choose between WiFi, Bluetooth, ZigBee, LoRaWan, and Z-Wave as their network protocol. Moreover, there is no standard solution for handling security issues in IoT, such as IPSec or SSL for the Internet. In the application layer, Constrained Application Protocol (CoAP) and MQTT are mainly used between the constrained IoT devices and servers. MQTT protocol \cite{MQTTIssues} has some limitations. For example, it can be used only for very low processor devices and can communicate mainly with Amazon cloud applications for servers. CoAP uses RESTful architecture to access the resources from a server through URI (Universal Resource Identifier) and message communication. \cite{restfulCoAP}, \cite{standardIoT} These two application layer protocols provide security features which are far from being perfectly efficient.
 \subsection{The Problem of AKE}
  One of the classic security problems is the design of authenticated key exchange protocols in which two parties authenticate each other and agree on a session key. The parties involved have to prove their claimed identity. The AKE protocol should be secure against an active adversary who can inject, modify, and delete packets. Most common AKE protocols rely on a trusted third party to certify the claimed identity of parties. In Transport Layer Security (TLS) handshake, for example, the server's identity is verified using the server's public key and the digital signature of the SSL certificate authority. In Kerberos, the Authentication Server and the Ticket Granting Server are trusted third parties. \cite{steiner1988kerberos} Some AKE protocols, such as Google's QUIC protocol, are multi-stage. In these protocols, different runs of the protocol are not independent, and the former keys are used to derive the next key \cite{fischlin2014multi}. In some cases, AKE is based on a pre-shared secret between parties. The high entropy pre-shared secret ensures forward secrecy. In this paper, we present LAKEE, an AKE protocol which is certificateless and based on a pre-shared secret. Our protocol avoids the public key infrastructure by using symmetric keys to be as lightweight as possible for employment in IoT networks. \cite{SAKE}
\par We implement our protocol as an application on top of CoAP. Research has an interest in building application layer protocols, such as HTTP \cite{HTTP-COAP} and SOAP \cite{SOAP}, on top of CoAP. Our practice is the same.
\par The contributions of the present work can be summarized below.
\begin{itemize}
    \item We propose a lightweight and effective protocol which guarantees secure communication by establishing a secure session. We have implemented our protocol in Python using the aiocoap \cite{aiocoAP} library.
    \item The protocol can easily integrate with the existing system deploying DTLS-PSK without requiring  additional code.
    \item After session establishment, the transport layer can use the generated key to encrypt application layer messages securely. More precisely, the session parameters are mapped to the DTLS session parameters without requiring any modification in the existing session format for DTLS-PSK mode. Thus the transport layer can protect the application layer message using the encryption mechanism of DTLS-PSK.
    \item We estimate our protocol's communication and computation overhead and compare it with the alternative protocols. We prove that the overhead of the protocol is significantly less than its competitors.
    \item We verify the security strength of the proposed protocol both by formal logic and simulation tools. 
\end{itemize}
\section{Related Work}
Many researchers have focused on balancing the tradeoff between security and overhead of session establishment protocols in the context of IoT devices. Pan \textit{et al.} \cite{Pan} offer a lightweight certificateless protocol based on elliptic curves and bilinear pairing. However, a third component, as a key generation center (KGC), is required. The total number of messages for key exchange is also 4, which is more than our proposed protocol. It is not clear which participant is responsible for initiating the protocol.
\par Kalra \textit{et al.} \cite{Kalra} and Kumari \textit{et al.} \cite{Kumari} proposed authentication methods, but they depend on HTTP cookies. Moreover, when the cookie of a user-server connection resets, the number of steps is increased to 5. It is also shown that the Kalra and Kumari protocols are not desirably secure. \cite{Kumari}, \cite{adesh}
\par Shin \textit{et al.} \cite{shin2016MQTT} have proposed a security framework for MQTT, enabling the client and server to come up with a symmetric session key based on the AugPAKE protocol. \cite{shin2012augpake}
The protocol proposed by Yeh \textit{et al.} \cite{YehWirelessECC} is a lightweight authentication method which employs elliptic curve cryptography and hashing to provide two endpoints with a session key. Hashing is known to be more efficient than encryption and can therefore benefit IoT networks. However, the method is dependent on a third element called Registration GW-node. Moreover, the login phase has 4 messages. 
Albalas \textit{et al.} apply both Rivest–Shamir–Adleman (RSA) and Elliptic Curve Cryptography (ECC) for key generation in CoAP and discuss the advantages of ECC. 
The session key establishment scheme proposed by Dey and Hossain \cite{dey2019session} has 7 steps involving 3 participants; the smart device, the service provider, and the home gateway.
\par Bhattacharyya \textit{et al.} \cite{less} have proposed LESS, a simple and lightweight authenticated key exchange protocol for IoT devices. The proposed protocol is implemented on top of the CoAP layer. The main difference between LESS and our protocol is the session key generation. In LESS, the server is responsible for choosing the session keys, which are transferred to the client in the second and fourth messages of the protocol. Majumder \textit{et al.} \cite{Majumder} have also proposed an authenticated key exchange protocol based on ECC. They verify the security strength of their protocol both formally (using the BAN logic) and practically (using the AVISPA simulation tool). However, their protocol, ECC-CoAP, is not implemented, and its compatibility with CoAP is not verified. For example, CoAP may unexpectedly fragment the messages if they are larger than a limit.  
\section{Background}
\subsection{Constrained Application Protocol (CoAP)}
\begin{figure}
	\centering
	\includegraphics[width=\linewidth]{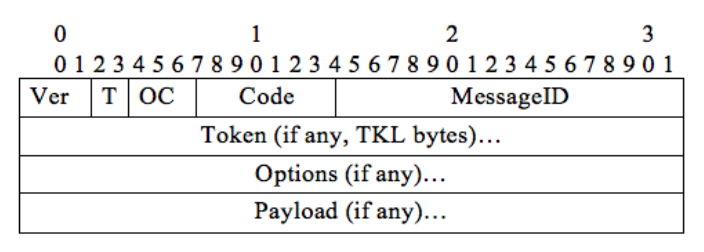}
	\caption{CoAP message structure \cite{alhaj2018constrained}}
	\label{fig:COAP-MSG}
\end{figure}

CoAP is an application layer protocol for resource-constrained internet devices. For web integration, CoAP can be converted to HTTP. It also meets the fundamental requirements of IoT communication, which are multicast support, low overhead, and simplicity. 
CoAP is built on top of the User Datagram Protocol (UDP). As UDP is unreliable, unlike Transmission Control Protocol (TCP), an optional request/response layer is included in the messaging. The CoAP message structure is shown in figure \ref{fig:COAP-MSG}
CoAP supports four types of messages: Confirmable (CON), Non-Confirmable (NON), Acknowledgement (ACK), and Reset (RST). A confirmable message guarantees reliability. Each confirmable message is replied to with an ACK by the server with the same message ID. Otherwise, it will be retransmitted using a default timeout and exponential back-off between retransmissions. However, non-confirmable messages are unreliable and are not replied to with an ACK. If the recipient cannot process a Non-confirmable message, it may reply with an RST message.

\subsection{Elliptic Curve Cryptography (ECC)}
ECC, proposed by Victor Miller and Neal Koblitz \cite{ECCMiller}, \cite{ECCKoblitz}, is a key-based cryptography technique based on the algebraic structure of elliptic curves over finite fields. It usually is an alternative to the RSA cryptographic algorithm. ECC features smaller ciphertext, keys, signatures and a faster generation of keys and signatures. The fundamental downside of ECC is that, due to the complexity of its mathematical foundation, it is difficult to implement securely.
\par An elliptic curve $E$ is defined over a prime finite field $F_p$. This smooth projective curve of genus 1 is denoted as $E/F_P$ and holds the following elliptic curve equation:
$$y^2 mod \, p = (x^3 + ax + b ) mod \, p$$
We employ the mathematical operation of \textbf{scalar point multiplication} to perform an Elliptic Curve Diffie–Hellman Key Exchange (ECDH). 
The security strength of \textit{ECC} lies in the difficulty of solving the \textit{Elliptic Curve Discrete Logarithm Problem (ECDLP)}. Similar to \textit{the Diffie-Helman Problem (DHP)}, \textit{ECDLP} is based on the discrete logarithm problem and does not pursue any polynomial-time algorithm.
\subsection{Datagram Transport Layer Security (DTLS)}
IoT applications frequently encounter a variety of communication patterns that the Transmission Control Protocol (TCP) is unable to support effectively. Due to energy constraints, devices may often go into sleep mode; thus, it is infeasible to maintain a long-lived connection in IoT applications. In consequence, IoT networks use UDP as the transport protocol. Therefore, the SSL/TLS protocol, which makes TCP secure, is not applicable to the IoT environment. IoT networks can benefit from DTLS, which is a derivation of TLS protocol and provides the same security services on top of UDP.
\par
\begin{figure}
	\centering
	\includegraphics[width=\linewidth]{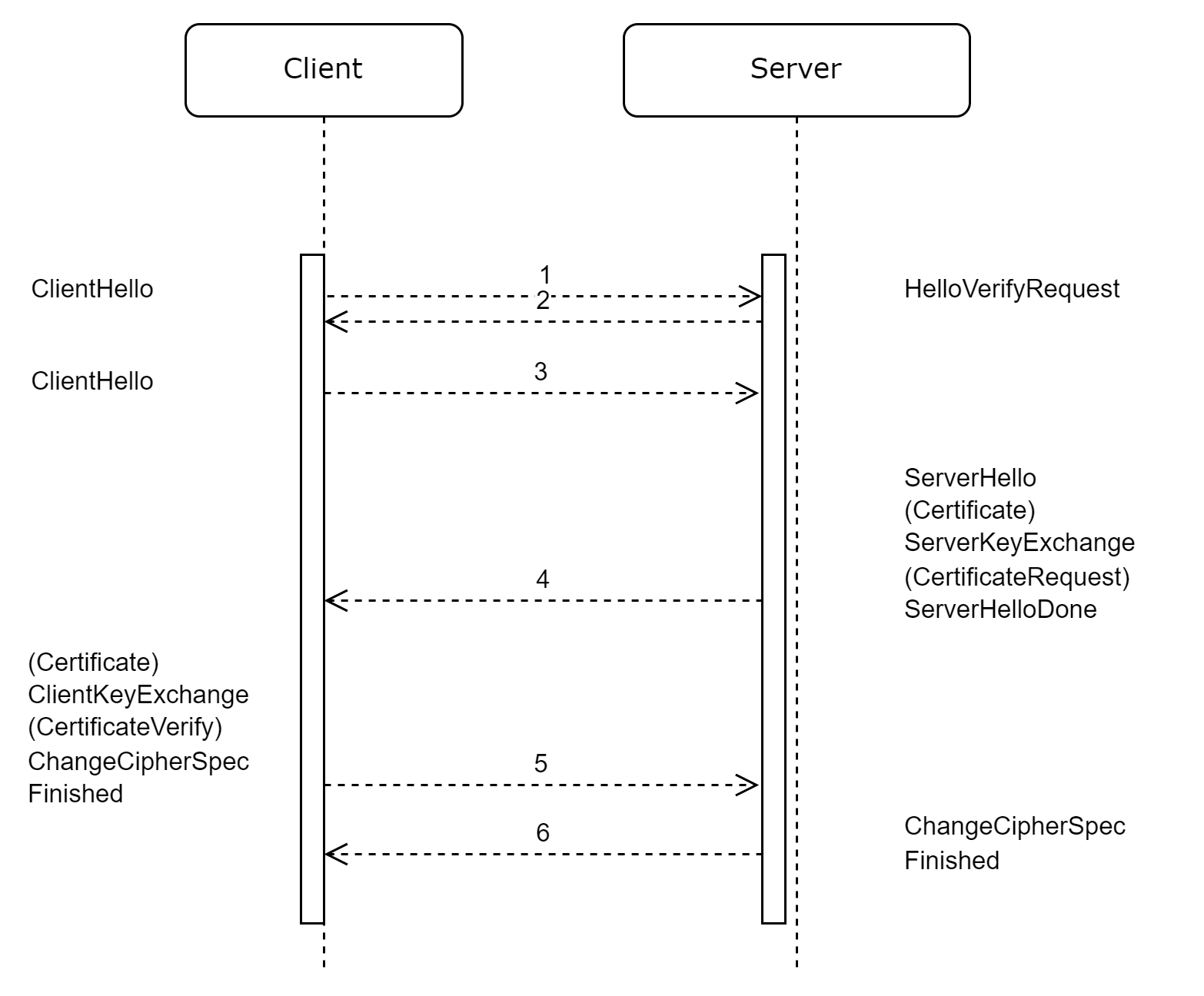}
	\caption{DTLS session-establishment in the PreSharedKey mode}
	\label{fig:DTLS_PSK}
\end{figure}

 CoAP can optionally use DTLS. CoAP defines four security modes.
\begin{itemize}
  \item NoSec.
  \item PreSharedKey,
  \item RawPublicKey,
  \item Certificate
\end{itemize}
DTLS provides CoAP with a nonoptimal handshake protocol with a long-lived key in the PresharedKey mode. It takes 6 messages (7 steps) to establish a session with this handshake. \cite{RFC4279}, \cite{RFC6347}
DTLS handshake protocol which can be used in CoAP for session establishment, is illustrated in figure \ref{fig:DTLS_PSK}.
\section{Proposed Protocol}
\begin{figure}
	\centering
	\includegraphics[width=\linewidth]{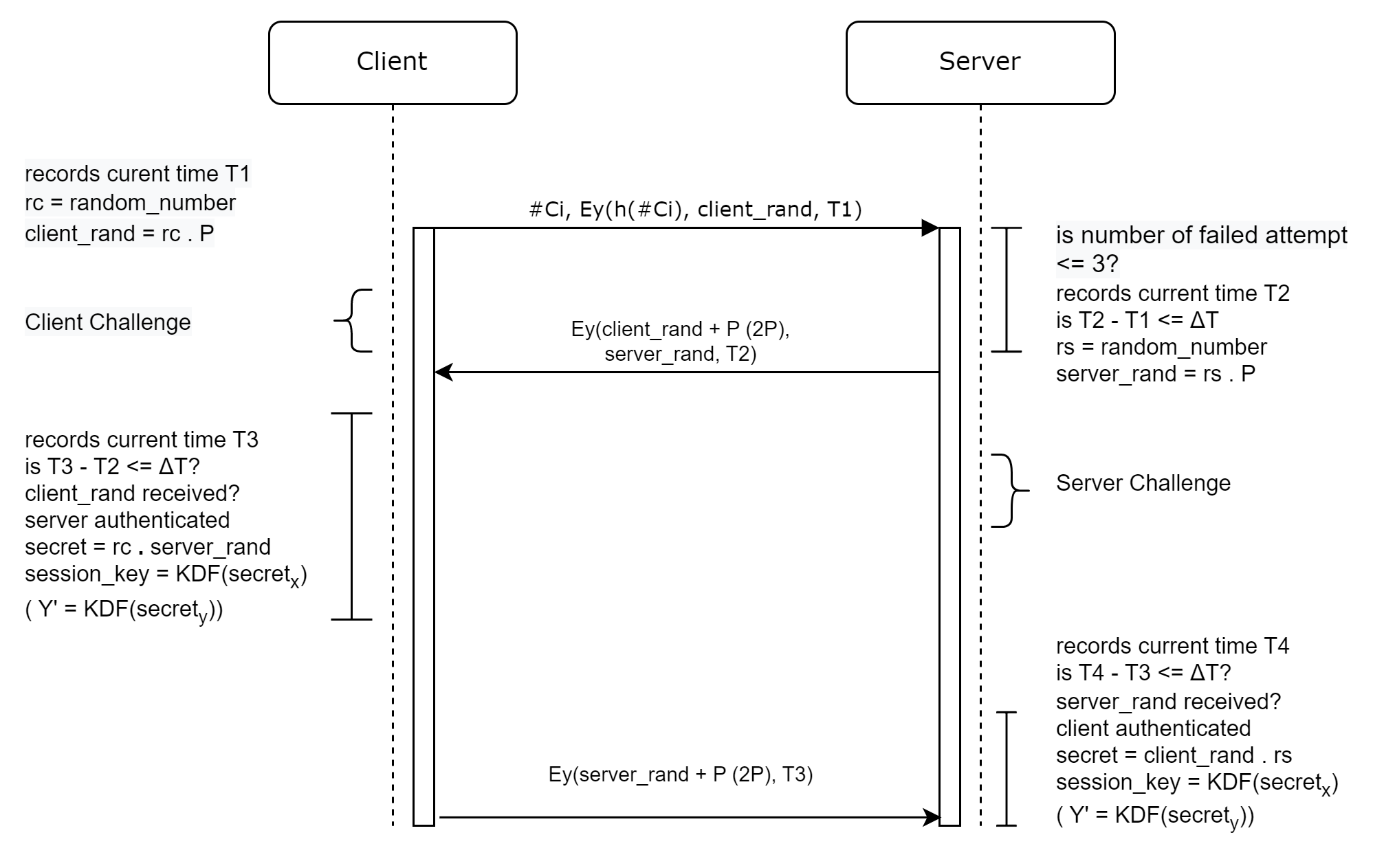}
	\caption{Proposed protocol for authenticated key exchange}
	\label{fig:Protocol}
\end{figure}

In this section, we present the LAKEE protocol completely. $X \rightarrow S: M$ denotes X sends message M to S. Figure \protect\ref{fig:Protocol} illustrates the protocol, which has 4 steps including 3 messages, while Table \rom{1} explains the notations used. 
\begin{table*}[htbp]
\caption{Symbols \& Notation table}
    \begin{center}
    \begin{tabular}{|c|c|}
        \hline
        \textbf{Notation}&\textbf{Description} \\
        \hline
        C & client/IoT device \\
        \hline
        S & server \\
        \hline
        $F_p$ & large prime finite field over \textit{p} \\
        \hline
        P & a generator point on $E/F_p(a,b)$ with order \textit{n} \\
        \hline
        $E_p(a,p)$ & an elliptic curve over the finite field $F_p$ and cyclic group generator $P$ \\
        \hline
        Y & long-term symmetric key between client and server \\
        \hline
        $r_c$ & random number selected by the client \\
        \hline
        $r_s$ & random number selected by the server \\
        \hline
        $client\_rand$ & client's shared data for producing the session key where $client\_rand=r_c.P$\\
        \hline
        $server\_rand$ & server's shared data for producing the session key where $server\_rand=r_s.P$\\
        \hline
        $[E(.)_k]$ & authenticated encryption of a plaintext using a symmetric key \textit{k} \\
        \hline
        $h(m)$ & unkeyed one-way hash value of message m \\
        \hline
        $\|$ & concatenation \\
        \hline
        $\#C_i$ & client identifier \\
        \hline
        $a.P$ & scalar point multiplication; Point P is multiplied by scalar a. \\
        \hline

        \end{tabular}
    \end{center}
\end{table*}

\subsection{Pre-negotiation}
Before starting the handshake, the client and server must have agreed upon a prime finite field, an elliptic curve, a generator point, a key derivation function, and the symmetric high-entropy  key Y. 

\subsection{Step 1. Client Challange}
$$C \rightarrow S: \#C_i, E_y(h(\#C_i), client\_rand, T1)$$
\par At first, the client generates a random number ($r_c$). It then computes client's challenge as $ client\_rand = r_c.P$. The client concatenates the hash of its ID, client\_rand, and current timestamp and encrypts the outcome with the pre-shared key Y. The client then sends its ID with the encrypted message to the server. The server recognizes the raw identifier, associating this message with a session initiation request. The client reveals his identity in this message. The novelty of our protocol is that the client challenges and authenticates the server first, reducing the required number of messages. 

\subsection{Step 2. Server Response and Challenge}
$$S \rightarrow C: E_y(client\_rand + P, server\_rand, T2)$$
\centerline{or}
$$S \rightarrow C: E_y(client\_rand + 2P, server\_rand, T2)$$
\par After receiving the session initiation message from the client, the server checks the number of failed session initiation requests within a time frame originating from the message's source IP address. If it's greater than 3, the message is not processed. Otherwise, the server tries to decrypt the message using the pre-shared key. The server records the current time and checks $|T_2-T_1|\leq \Delta T$. If it holds, the shared secret is retrieved, and a random number is generated ($r_s$). The server then computes the server's challenge as $ server\_rand = r_s.P$. It decides whether it is time to change the long-term secret (Y). The strategy with which the server chooses the time is out of the scope of the paper and is dependent on the information it stores about the clients. If the server decides to maintain Y, it sums the client's challenge with generator P ($client\_rand$ is a point on the curve). It then sends the result alongside its own challenge to the client. If it decides to change Y during the handshake, it sums the client's challenge with twice the generator (2P) and sends the result alongside its own challenge to the client. When the information in the incoming is irrelevant, the handshake is terminated, and the number of failed attempts from the client's IP range is increased.

\subsection{Step 3. Server Authentication and Client Response}
$$C \rightarrow S: E_y(server\_rand + P,T3)$$
\centerline{or}
$$C \rightarrow S: E_y(server\_rand + 2P,T3)$$
\par In this stage, the client decrypts the server's message with the shared secret. It records the current time and checks $|T_3-T_2|\leq \Delta T$. If it holds, the client verifies the server's response provided that the message contains $client\_rand + P$ or $client\_rand + 2P$. In the second case, it realizes that it should update Y during this handshake. The client then calculates $server\_rand + P$ or $server\_rand + 2P$ accordingly. It concatenates it with the current timestamp, encrypts the outcome, and sends the message to the server. The client locally calculates a shared secret which is a point on the elliptic curve. 
$$secret = rc.server\_rand$$
Using the horizontal position of the secret on the curve and the agreed key derivation function (KDF), the client calculates the session key. 
$$session\_key = KDF(secret_x)$$
If it also had to update Y, it calculates a new long-term key using the vertical position of the secret on the curve and the KDF.
$$Y'= KDF(secret_y)$$
\subsection{Step 4. Client Authentication}
\par In this stage, the server decrypts the client's message with the shared secret. It records the current time and checks $|T_4-T_3|\leq \Delta T$. If it holds, the server verifies the client's response the same as the previous step and calculates the session key.
$$secret = rs.client\_rand$$
Using the horizontal position of the secret on the curve and the KDF, the server calculates the session key. 
$$session\_key = KDF(secret_x)$$
If it also had to update Y, it calculates the following.
$$Y'= KDF(secret_y)$$
\par In contrast with typical Internet users, each IoT device reaches out to a limited number of servers. We assume, as a consequence, that providing each client-server pair with a shared secret is feasible. Our certificateless protocol is independent of any authorities and third parties. Therefore, the proof of the parties' identity relies on the possession of the pre-shared and symmetric secret. As it is certificateless, our protocol with minor changes is employable in mesh IoT networks where any node can act as a client or a server dynamically. 

\section{Implementation}
We have implemented our protocol on top of CoAP using the \textit{aiocoap} python library. \cite{aiocoAP} We aimed to have a reproducible implementation, and it serves only as self-documenting code. We acknowledge that it could be more efficient by employing C++, for example. It is publicly available on Github\footnote{Please refer to \url{https://github.com/sina-nabavi/LAKEE-over-aiocoap}.}. The client performs all of the steps sequentially and in a single function. In contrast, the server would not block for an individual client. Therefore, it must store the handshake progress information, such as the challenge it sent to that particular client, and process the corresponding client's response upon reception. The first message (client challenge) is a \textbf{Confirmable} (CON) message. The second message (server response and challenge) is the \textbf{Acknowledgement} (ACK) of the first, and the third message's (client response) type is \textbf{Non-Confirmable} (NON). After sending the third message, the client would not wait for a reply from the server.
\par We have  used the \textbf{PyCryptodome} package for encryption and elliptic curve operations. The AES block cipher is used because it has been standardized and resistant to side-channel attacks. Moreover, there are hardware acceleration mechanisms available for AES. The cipher in the CCM mode calculates a message authentication code (MAC) for each message to verify whether the received data is authentic.  We employ the curve \textit{edwards448}, also known as \textit{Ed448-Goldilocks}, with the following equation. \cite{hamburg2015ed448}
$$-x^2 + y^2 = 1 - 39081x^2y^2$$
The curve aims for a 224-bit security level, meaning that it is designed to increase the cost of breaking a discrete logarithm computation to $2^{224}$ bit operations. \cite{RFC7748} Daniel J. Bernstein and Tanja Lange have shown that this curve meets the ECDLP and ECC requirements. \cite{SafeCurves} However, \textit{Edwards448} points and scalars are bigger than those of other curves.
We use a fixed 128-bit shared secret and the base point of the curve as the generator. \textit{PBKDF2} is used with \textit{SHA512} for deriving the session keys and new long-term secrets. \cite{PBKDF2RFC2898} The ECDH \textit{secret} has enough entropy for us to derive the keys without a salt and a large iteration number. The iteration number is 16 in the implementation.  
\par The code reports the execution time at the end of the handshake. It varies by the running hardware. Once the key exchange has finished, the responsibility can shift from CoAP to DTLS for providing channel security. With the correct parameters, a DTLS session can be instantiated without having to perform the DTLS\_PSK 7-step handshake.
	\begin{itemize} 
	\item cipher suite = TLS\_PSK\_WITH\_AES 128 CCM 8
	\item read state = S\_k.x with encryption nonce
	\item write state = S\_k.y with encryption nonce
	\end{itemize}

Along with the above parameters, one has to instantiate the client and server's initialization vectors, generate a sequence number, and fill other parameters for the desired DTLS session, such as Peer certificate with NULL.

\section{Security Evaluation}
In this section, we evaluate the security strength of our protocol.
\subsection{Properties of the Protocol}
\par 
The security goals are achieved as stated below.
\begin{itemize} 
\item \textbf{Confidentiality:}
All the secrets, including the timestamps and challenges, are encrypted. 
\item \textbf{Integrity:}
We use the AES-CCM ciphersuite, which performs MAC(message authentication code)-then-Encrypt for every message. The MACs and the hash of the client ID guarantee the integrity of messages.
\item \textbf{Entity Authentication:}
Instead of digital signatures and certificates, the approach of this protocol for entity authentication is challenge-response. The random numbers used for building the session keys are used as challenges, and the other party needs the pre-shared secret and the negotiated curve to be able to process them.
\item \textbf{Message Authentication:}
As stated, MACs computed for transmitted messages provide message authentication.
\item \textbf{Non-repudation:} 
As the client is expected to be an IoT device for which there are pre-defined tasks, we do \textbf{not} count non-repudiation as a necessary property for our protocol. Hence, we base our protocol on the pre-shared \textbf{symmetric} key. Also, instead of digital signatures, which provide non-repudiation, we use MACs which provide authentication.

\item \textbf{Perfect Forward Secrecy:} 
In the proposed scheme, if the key Y is compromised and the adversary is able to read $client\_rand$ and $server\_rand$, he cannot calculate the session key as he should obtain at least one of the secret randoms. The adversary cannot obtain the secret randoms from $client\_rand$ and $server\_rand$ due to the hardness of the elliptic curve discrete logarithm problem.
\end{itemize}

\subsection{Quantitative Security Analysis}
In this part, we discuss the security of our protocol against famous attacks. Table
\ref{table:sec-comp} compares the security  strength of LAKEE with alternative protocols.

\subsubsection{Man-in-the-Middle Attack}
The assumption of our scheme is that the pre-shared key (i.e., $Y$) remains a secret. Consequently, an adversary would not be able to modify plain messages, including the session initiation message. Besides, the MACs computed for each message make them unforgeable.
\subsubsection{Denial-of-Service (DoS) Attack} 
The number of times an IoT device can attempt to be authenticated is limited within a timeframe. In the session initiation phase of our scheme, if the IoT device fails to be authenticated by the server within 3 attempts, the IoT device will be blocked for a specific period. Hence, an adversary will not be able to send multiple requests (more than 3) to overload the system resource to make the services unavailable to the legitimate client. Thus, our protocol restricts the DoS attack.

\subsubsection{Replay Attack} 
A timestamp is associated with each message. If a participant receives an old message, it is not processed, and the connection is terminated. Hence, an adversary may not be able to use the captured messages to perform a replay attack. The adversary is also not able to change the timestamps as they are encrypted and protected with MAC.

\subsubsection{Insider Attack} 
In this type of attack, an insider of the remote server acts as an adversary after getting some user credentials stored on the remote server. As the session key cannot be computed by acquiring $client\_rand$ and $\_rand$ and the adversary needs to know at least one of the random secrets, the protocol is safe against insider attack.

\subsubsection{User and Server Impersonation Attack} 
As the attacker is not in possession of key Y, it cannot impersonate either client or server as all the messages include MACs and cannot be forged.

\subsubsection{Offline Password Guessing Attack} 
This attack mainly occurs in password-based authentication schemes, which is also a challenge-response scheme. However, in our proposed scheme, the challenge will differ at each epoch. Hence, the attacker cannot randomly guess the response to authenticate himself.

\subsubsection{Known Session Specific Temporary Attack} 
The session key is calculated at the ends, not transmitted with the messages. Hence, LAKEE is secure against known session specific temporary attack.

\subsubsection{Invalid Curve Attack}
When the attacker forces the server to compute a common secret using a point outside the defined curve, an invalid curve attack has occurred. As the client and server verify random points they receive, i.e., $client\_rand$ and $server\_rand$, they would terminate the connection if each point is in infinity or outside the chosen curve.
\subsubsection{The MOV attack} 
As all the points are transferred encrypted, we would not expect an adversary to obtain the random points (i.e., $server\_rand, client\_rand$). If any of the random points are guessed or acquired in polynomial time, there are two ways to prevent the adversary from obtaining the generator point (i.e., \textit{P}). 1) Keep the embedding degree \textit{k} greater than 100. 2) If the curve's subgroup divides the curve order, then the embedding degree should be the smallest $k \ge 2$ such that $n | q^k - 1$ for all values \textit{k} from 2 to 100.

\begin{table*}[htbp]
\caption{Security Strength}
    \begin{center}
    \begin{tabular}{|c|c||c||c||c|}
        \hline
        \textbf{Secure Against}& LESS \cite{less} & Dey \textit{et al.} \cite{dey2019session} & ECC-CoAP \cite{Majumder}  & LAKEE\\
        \hline
        Man-in-the-middle attack & Yes & Yes & Yes & Yes \\
        \hline
        Denial-of-service (DoS) attack & No & No & Yes & Yes \\
        \hline
        Replay attack & Yes & Yes & Yes & Yes \\
        \hline
        Insider attack & Yes & Yes & Yes & Yes \\
        \hline
        User impersonation attack & Yes & Yes & Yes & Yes \\
        \hline
        Server impersonation attack & Yes & No & Yes & Yes \\
        \hline
        Offline password guessing attack & No & No & Yes & Yes \\
        \hline
        Known session specific temporary attack & Yes & Yes & Yes & Yes \\
        \hline
        Session key computation attack & Yes & Yes & Yes & Yes \\
        \hline
        Invalid curve attack & N/A & N/A & No & Yes \\
        \hline
        The MOV attack & N/A & N/A & No & Yes \\
        \hline
        \end{tabular}
    \end{center}
    
    \label{table:sec-comp}
\end{table*}

\subsection{AVISPA Tool}
In this section, we demonstrate that LAKEE is secure against relevant threats using the Automated Verification Internet Security Protocol and Applications (AVISPA) simulator program.
\subsubsection{Introduction to the AVISPA Simulation Tool}
\par
\begin{figure}
	\centering
	\includegraphics[width=\linewidth]{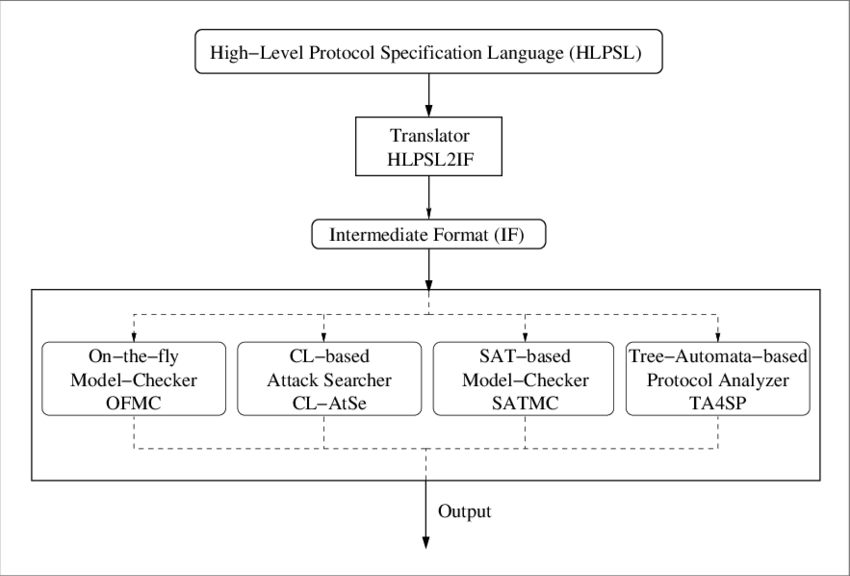}
	\caption{The architecture of the AVISPA tool}
	\label{fig:avispa}
\end{figure}

 AVISPA is a role-based simulator that assigns  a specific role to each protocol participant. A specification provided in a supported language is used to verify a protocol. The primary language supported by AVISPA is High-Level Protocol Specification Language (HLPSL). The working structure of AVISPA is depicted in figure \ref{fig:avispa}. HLPSL2IF translates HLPSL specification into an Intermediate Format (IF), where IF is a low-level language. The backends of the AVISPA tool use IF to analyze whether the security and authentication goals written in the specification are satisfied or violated. The AVISPA then produces the output SAFE and UNSAFE, respectively. The current version of the AVISPA tool supports 4 different types of backends. 
\begin{itemize}
    \item On-the-fly Model Checker (OFMC)
    \item Constraint logic-based attack searcher (CL-At Se)
    \item State of the Art based Model Checker (SATMC)
    \item Tree Automata based protocol for the security protocol analysis (TA4SP)
\end{itemize}

\subsubsection{OFMC}
The AVISPA tool includes several backends for analyzing protocols. We use OFMC, a tool for falsification or verification for a bounded number of sessions. OFMC employs the Dolev-Yao intruder model \cite{dolev-yao} to check several parallel sessions executed by honest agents. The input to this backend is a description of a transition system, and the output is either \textbf{SAFE}, showing that limited sessions are secure, or \textbf{UNSAFE} with an attack trace. 
\subsubsection{HLPSL}
We now discuss the basics of the AVISPA tool's protocol specification language. HLPSL is a role-based language with which we can specify the actions of each participant in a protocol. For specifying a protocol in the HLPSL language, we have to define the following.
\begin{itemize}
    \item \textbf{Basic Roles}: For both the client and the server, we have to define a role, including their actions, such as receiving and sending messages in certain formats.
    \item \textbf{Transitions}: Each transition is the representation of the receipt of a message and the sending of a reply message. Transitions are defined for a certain role.
    \item \textbf{Composed Roles}: For grouping our basic roles and representing a whole protocol session, we have to define a \textit{session} role. For storing the global constants and composition of multiple sessions, we have to define an \textit{environment} role. The intruder can actively take action in this environment. These composed roles do not have a transition section, but they have a composition section in which the basic roles are instantiated.
    \item \textbf{Security Goals}: In the environment role, we have to declare which parameters are shared secrets and which parameters are used for the authentication of participants.
\end{itemize}
\par For more information about the syntax of HLPSL and examples, please refer to the AVISPA manual. \cite{team2006avispa}
\subsubsection{Simulation Results}
We have defined our protocol in HLPSL with the following simplifications. We have not included timestamps and have defined elliptic curve mathematical functions simply as hash functions.
\begin{figure}
	\centering
	\includegraphics[width=\linewidth]{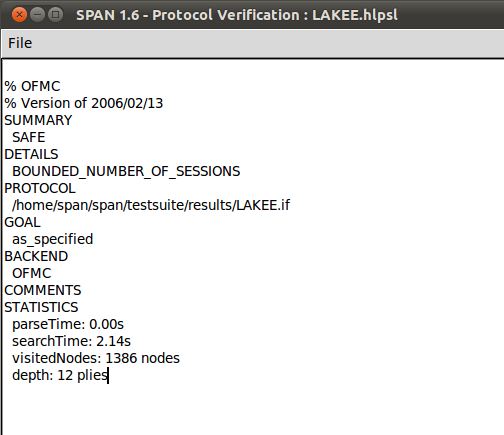}
	\caption{AVISPA verification result}
	\label{fig:avispares}
\end{figure}
\begin{lstlisting}
role client(C: agent, S:agent, K:symmetric_key,Mul: hash_func,
            X:text,
            Succ: hash_func,
            SND, RCV: channel(dy))

played_by C def=
    local State:nat,
    Ci: text,
    Rc, Client_rand, Server_rand, SK: message
    
    const
    sec_1, sec_2, sec_3, sec_4,
    auth_1,auth_2:protocol_id
    
    init State:= 0
    
    transition
        
        1. State=0 /\ RCV(start) =|>
            State':=2 /\ Rc' := new()
                        /\ secret(Rc', sec_1,{C})
                        /\ Client_rand' := Mul(Rc'.X)
                        /\ Ci' := new()
                        /\ SND({Succ(Ci').Client_rand'}_K)
                        /\ secret(Client_rand', sec_2,{C,S})
                        
        2.    State=2 /\ RCV({Succ(Client_rand).Server_rand'}_K)=|>
            State':=4 /\ SK':=Mul(Rc.Server_rand')
                        /\ SND({Succ(Server_rand')}_K)
                        /\ witness(C,S,auth_2, Server_rand')
                        /\ request(C,S, auth_1, Client_rand)
end role

%%%

role server(S: agent,C:agent, K: symmetric_key, Mul: hash_func,
            X: text,
            Succ: hash_func,
            SND,RCV:channel(dy))

played_by S def=
    local State: nat,
    Ci: text,
    Rs, Server_rand, Client_rand, SK: message
    
    init State:=1
    transition
        
        1. State=1 /\ RCV({Succ(Ci').Client_rand'}_K)=|>
            State' :=3 /\ Rs' := new()
                        /\ secret(Rs', sec_3,{S})
                        /\ Server_rand' := Mul(Rs'.X)
                        /\ secret(Server_rand', sec_4,{S,C})
                        /\ SND({Succ(Client_rand').Server_rand'}_K)
                        /\ witness(S,C, auth_1, Client_rand')
                        
        2. State=3 /\ RCV({Succ(Server_rand)}_K)=|>
            State' := 5 /\ request(S,C, auth_2, Server_rand)
                        /\ SK' := Mul(Client_rand.Rs)
                        
end role

%%%

role session(C: agent, S: agent, K: symmetric_key, Mul: hash_func, X: text, Succ: hash_func)
def= local SNDC, RCVC, SNDS, RCVS: channel(dy)
    composition
        client(C, S, K,  Mul, X, Succ, SNDC, RCVC) /\
        server(S, C, K, Mul, X, Succ, SNDS, RCVS)
end role

%%%

role environment()
def=
    const
        c,s, i: agent,
        kcs, kci, kis: symmetric_key,
        x: text,
        succ: hash_func,
        mul: hash_func
    
    intruder_knowledge = {c,s, kci, kis, succ, mul}
    composition
        session(c,s,kcs,mul,x,succ) /\
        session(c,s,kcs,mul,x,succ) /\
        session(c,i,kci,mul,x,succ) /\
        session(i,s,kis,mul,x,succ)
    end role
    goal
        secrecy_of sec_1
        secrecy_of sec_2
        secrecy_of sec_3
        secrecy_of sec_4
        authentication_on auth_1
        authentication_on auth_2
    end goal

environment()
\end{lstlisting}
In our specification, the client authenticates the server on comprehension of \textit{SRand} (server\_rand), and the server authenticates the client on comprehension of \textit{CRand} (client\_rand). \textit{SRand, CRand, Srand (rs), and Crand (rc)} are shared secrets. We have not included the session key in shared secrets as it is not transferred in the protocol.
We have used the SPAN + AVISPA virtual machine and SPAN graphical tool \cite{heen2008industrialspan} to parse our specification and the OFMC backend to verify our protocol. The result in figure \ref{fig:avispares} indicates that our protocol is safe.
\subsection{Formal Verification Using the SVO Logic}
Although Majumder \protect\textit{et al.} \cite{Majumder} and Mahmood \textit{et al.} \cite{Mahmood} use the BAN logic \cite{Burrows} in order to verify their protocol formally, we firmly believe that BAN logic is not suitable for verifying key exchange protocols. Alternatively, we use the SVO Logic \cite{syverson1996unified}. 
Syverson and Oorschot \cite{syverson1996unified} revealed that BAN analysis might lead to inappropriate conclusions in some settings. Therefore, they presented a logic capturing the desirable properties of the BAN family of logics, including BAN \cite{Burrows}, AT \cite{abadi1991semantics}, GNY \cite{gong1990reasoning}, and VO, \cite{oorschotVO} that is both sound and relatively easy to use
\subsubsection{\textbf{Preliminaries of SVO Logic}}
In this section, we briefly review the SVO logic, the notations, rules, and axioms. However, we omit some definitions unused in our analysis, such as the concept of public key. \cite{TheLogicofAuthenProt}
\paragraph{\textbf{Notation}}
The entities' names who want to perform an authenticated key exchange are \textit{Client} and \textit{Server}. We have assumed that the Client and Server have shared a long-term key. We use the symbols for Client (C), Server (S), the pre-shared key (Y), and the session key (k).
\par P \textbf{\textit{believes}} X : The principal P may act as though X is true.
\par P \textbf{\textit{received}} X : The principal P received a message containing X. P can read and repeat X.
\par P \textbf{\textit{said}} X : The principal P at some time sent a message including X.
\par P \textbf{\textit{says}} X : The principal P must have said X since the beginning of the current protocol run.
\par P \textbf{\textit{has}} X : X is either,
\begin{itemize}
    \item Initially available to P,
    \item Received by P,
    \item Freshly generated by P,
    \item Constructible by P from the above.
\end{itemize}
\par P \textbf{\textit{controls}} X : The principal P has jurisdiction over X. P is an authority on X and should be trusted on this matter.
\par \textbf{\textit{fresh}} (X) : X has not been sent in any message prior to the current protocol run.
\par $P \xleftrightarrow{\text{k}} Q$ : P and Q may use the shared key k to communicate. k will never be discovered by any principal but P,Q, or a principal trusted by P or Q.
\par $PK_{\delta}(P, k)$ :  k is a public key-agreement key of P. A Diffie-Hellman key formed with k is shared with P.
\par $\{M\}_k$ :  Encryption of message M using key k. Encrypted messages are uniquely readable and verifiable as such by holders of the right keys. 
\par $P \xleftrightarrow{\text{k}{-}} Q \equiv (P \xleftrightarrow{\text{k}} Q \wedge P \text{ has } k)$ : No one aside from P and Q and those they trust knows or could deduce k. While P knows k, Q may or may not. k is said to be P's \textit{unconfirmed secret}.
\par $P \xleftrightarrow{\text{k}{+}} Q$ : P knows k, and has received evidence confirming that Q knows k. No parties other than P and Q and those they trust knows k. k is P's confirmed secret suitable for Q.
\paragraph{\textbf{SVO Rules}}
The SVO logic has two inference rules:
\par \textbf{Modus Ponens} : From $\varphi \,$  and $\varphi \rightarrow \psi$ infer $\psi $
\par \textbf{Necessiation} : From $\vdash \varphi $ infer $\vdash\text{P believes} \, \varphi$
\par '$\vdash$' is a metalinguistic symbol. '$\Gamma \, \vdash \, \varphi$' means that $\varphi$ is derivable from the set of formulae $\Gamma$ and the stated axioms using the inference rules. '$\vdash \varphi$' is a theorem, i.e., derivable from axioms alone without any additional assumptions. 
\paragraph{\textbf{SVO axioms}}
In this part we introduce the SVO axioms.
\newline
\newline \textbf{Belief Axioms}
\newline 1. $(\textit{ P believes } \varphi \wedge \textit{ P believes } (\varphi \rightarrow \psi)) \rightarrow \textit{ P believes } \psi$
\newline 2. $\textit{ P believes } \varphi \rightarrow \textit{ P believes (P believes } \varphi )$
\newline
\newline\textbf{Source Association Axiom}
\newline 3. $(P \xleftrightarrow{\text{k}} Q \wedge \textit{R received } \{ \textit{X from Q} \}_k ) \rightarrow (\textit{Q said X} \wedge \textit{ Q has X})$
\newline
\newline\textbf{Key Agreement Axiom}
\newline 4. $(PK_\delta(P, k_P) \wedge PK_\delta(Q, k_Q)) \rightarrow P \xleftrightarrow{F_{0(k_P,k_Q)}} Q$
\par $F_0(k,k')$ implicitly names the (Diffie-Hellman) function that combines $k$ with $k^-1$ (or, $k'$ with $k^-1$) to form a shared key. This is the axiom the BAN logic lacks.
\newline
\newline\textbf{Receiving Axioms}
\newline 5. $\textit{P received }(X_1,...,X_n) \rightarrow \textit{P received } X_i, \textit{for i = 1,...,n}$
\newline 6. $(\textit{P received} \{X\}_{k+} \wedge \textit{P has }k ^ -) \rightarrow \textit{P received X}$
\par Here $K ^ +$ and $k ^ - $ are used to abstractly represent cognate keys, wheter symmetric or assymetric. In the symmetric case, $k ^ + = k ^ - = k$. In the assymetric case, $k^+$ is a public key and $k^-$ is the private key.
\newline
\newline\textbf{Posession Axioms}
\newline 7. $\textit{P received X } \rightarrow \textit{P has X}$
\newline 8. $\textit{P has } (X_1, ..., X_n) \rightarrow \textit{P has } X_i, \textit{for i = 1,...,n}$
\newline 9. $(\textit{P has } X_1 \wedge \textit{...} \wedge \textit{P has } X_n) \rightarrow
 \textit{P has } F(X_1,...,X_n)$
\par F is a meta-notation for any function computable in practice by P, e.g., encryption.
\newline
\newline\textbf{Comprehension Axiom}
\newline 10. $\textit{P believes} ( \textit{P has } F(X))\rightarrow \textit{P believes} ( \textit{P has X})$
\par F is meta-notation for any function that is effectively one-one and such that $F^+$ or $F^-$ is computable in practice by $P$.
\newline
\newline\textbf{Saying Axioms}
\newline 11. $\textit{P said }(X_1,...,X_n)\rightarrow \textit{P said } X_i \wedge  \textit{P has } X_i, \\ \textit{    for i = 1,...,n}$
\newline 12. $\textit{P says }(X_1,...,X_n) \rightarrow (\textit{P said} (X_1,...,X_n) \wedge \\ \textit{    P says} X_i), \textit{ for i = 1,...,n}$
\newline
\newline\textbf{Freshness Axioms}
\newline 13. $fresh(X_i) \rightarrow fresh(X_1,...,X_n), \textit{for i = 1,...,n}.$
\newline 14. $fresh(X_1,...,X_n) \rightarrow fresh \, F(X_1,...,X_n)$
\par F must genuinely depend on all component arguments. This means that it is infeasible to compute the value of F without value of all the $X_i$.
\newline
\newline\textbf{Jurisdiction and Nonce-verification Axioms}
\newline 15. $(\textit{P controls } \varphi \wedge \textit{P says } \varphi ) \rightarrow \varphi$
\newline 16. $(\textit{fresh(X) } \wedge \textit{P said X}) \rightarrow \textit{P says X}$
\newline
\newline\textbf{Symmetric Goodness Axiom}
\newline 17.  $P \xleftrightarrow{\text{k}} Q \equiv Q \xleftrightarrow{\text{k}} P$
\newline
\newline\textbf{Seeing Axioms}
\newline 18. $\textit{P received X } \rightarrow \textit{ P sees X}$
\newline 19. $\textit{P sees } (X_1, X_2, ..., X_n)\rightarrow \textit{ P sees } X_i$
\newline 20. $(\textit{P sees } X_1 \wedge ... \wedge \textit{ P sees } X_n)\rightarrow \\ \textit{ (P sees } F(X_1,...,X_n)) $
\newline
\subsubsection{\textbf{Verification of the Protocol using the SVO Logic}}
\par
We should meet the goals below to have a secure and authenticated key exchange.
\newline \textbf{G1. Entity authentication:} \textit{C believes S says $F(X, N_A)$}. This goal necessitates that, given $N_A$ known to be fresh to C, S recently sent a message $F(X, N_A)$, indicating that S has seen $N_A$ and has processed it.
\newline \textbf{G2. Secure key establishment:} Participants believe they are in possession of a good key to communicate with each other. The goal can be formulated as: $\textit{C believes C }\xleftrightarrow{\textit{k}{-}} \textit{ S} $
\par It is important to note that in our goals, the session key is an \textit{unconfirmed secret}. No one knows or could derive k except C and S and those they trust. This notation is different from \textit{confirmed secret}. If k is a \textit{confirmed secret}, C knows k and has received evidence that S also knows k. It is apparent that our protocol does not establish a confirmed secret.
\paragraph{\textbf{Assumptions}}
We need the following assumptions for deriving our goals. The assumptions hold for a successful handshake and not for a terminated one.
\newline
\textbf{P1.} $\text{C believes S } \xleftrightarrow{\text{Y}} \text{ C, S believes S }\xleftrightarrow{\text{Y}} \text{ C}$
\newline 
\textbf{P2.} $\text{C believes C sees }(client\_rand, server\_rand)$
\newline $\text{ S believes S sees }(client\_rand, server\_rand)$
\newline \textbf{P3.} $\text{C believes } \textit{fresh }(client\_rand, T_1,T_2,T_3)$
\newline $\text{ S believes } fresh(server\_rand, T_1,T_2,T_3)$
\newline \textbf{P4.} $\text{S believes S received } C_i,\{h(C_i), *_c, T_1\}_Y$
\newline \textbf{P5.} $\text{C believes C received } \{client\_rand + P, *_s, T_2\}_Y$
\newline \textbf{P6.} $\text{S believes S received } \{server\_rand + P, T_3\}_Y$
\newline \textbf{P7.} $\text{C believes }PK_\delta(C, client\_rand)$
\newline $\text{ S believes} PK_\delta(S, server\_rand)$
\newline \textbf{P8.} $\text{S belives (C says }(*_c) \rightarrow \text{C says }PK_\delta(C,*_c))$
\newline $\text{ C belives (S says }(*_s) \rightarrow \text{S says }PK_\delta(S,*_s))$
\newline \textbf{P9.} $\text{S believes C controls }PK_\delta(C,*_c)$
\newline $\text{ C believes S controls }PK_\delta(S,*_s)$
\newline
\paragraph{\textbf{Derivations}}
Based on the axioms, rules, and the assumptions, we can derive the following about the protocol.
\newline
$\textbf{1. } \text{S believes C said (server\_rand + P, T3)}$
\\\text{ by P6, P1, Axiom 3, Nec. rule}
\\ $\textbf{2. } \text{S believes fresh (server\_rand + P, T3)}$
\\\text{ by P3, Axiom 13, Nec. rule, Axiom 1, Modus Ponens (MP) rule}
\newline
$\textbf{3. } \text{S believes C says (server\_rand + P, T3)}$
\\ \text{ by 1, 2, Axiom 1, Axiom 16, MP rule, Nec. rule}
\\ $\textbf{4. } \text{S believes C says }F(N_s)$
\\ \text{ by 3, Axiom 12}
\par Derivation 4 satisfies our goal \textbf{G1} for authenticating C. By following the same steps for $client\_rand$, authentication of S is provable.
\newline
$\textbf{5. } \text{S believes C said }(C_i, h(C_i), *_c, T_1)$
\text{ by P7, P1, Axiom 3, Nec. rule}
\\ $\textbf{6. } \text{S believes fresh }(C_i, h(C_i), *_c, T_1)$
\\ \text{ by P3, Axiom 1, Axiom 13, Nec. rule, MP rule}
\\ $\textbf{7. } \text{S believes C says }(C_i, h(C_i), *_c, T_1)$
\\ \text{ by 5, 6, Axiom 1, Axiom 16, Nec. rule, MP rule}
$\textbf{8. } \text{S believes C says }(*_c)$ \newline
\text{ by 7, Axiom 12, Axiom 1, Nec. rule., MP rule}
\\ $\textbf{9. } \text{S believes C says }PK_\delta(C,*_c))$
\newline \text{ by 8, P8, Axiom 1, MP rule}
\newline $\textbf{10. } \text{S believes }PK_\delta(C,*_c))$
\newline \text{ by 9, P9, Axiom 1, Axiom 15, Nec rule, MP rule}
\newline $\textbf{11. } \text{S believes } S \xleftrightarrow{\text{k}} C $
\newline \text{ by 10, P7, Axiom 1, Axiom 4, Nec rule, MP rule}
\newline where k = $(*_c, server\_rand)$
\newline $\textbf{12. } \text{S believes S sees } (*_c)$
\newline \text{ by P4, Axiom 1, Axiom 18, Axiom 19, Axiom 20, Nec rule } 
\newline ,MP rule
\newline $\textbf{13. } \text{S believes S sees K}$
\newline \text{ by 12, P2, Axiom 1, Axiom 19, Axiom 20, Nec rule, MP }
\newline rule where k = $(*_c, server\_rand)$
\newline $\textbf{14. } \text{S believes } S \xleftrightarrow{\text{k}{-}} C $
\newline \text{ by 11, 13, Axiom 1, MP rule}
\par Derivation 14 satisfies our goal \textbf{G2}. By following the same steps, we can infer the key goodness goal for C as well.
\section{Performance Analysis}
In this section, we estimate the overhead of our protocol with respect to the arithmetic and cryptographic operations used in our scheme.
\subsection{Computation Overhead}
To calculate the total computational cost of our protocol, we have borrowed the primitive arithmetic and cryptographic operation timings from Kilinc \textit{et al.} \cite{Kilinc}, demonstrated in Table \protect\ref{table:crypto-time}. The table shows the arithmetic mean and the standard deviation of the following primitive operations for 1000 executions each.
\begin{table*}[htbp]
\caption{Execution time of different cryptographic operation}
    \begin{center}
    \begin{tabular}{|c|c|}
        \hline
        \textbf{Notation}&\textbf{Description and execution time (ms)} \\
        \hline
        $T_{ECPM}$ & Time complexity for the execution of the elliptic curve (ECC) point multiplication, $2.226 ms$ \\
        \hline
        $T_{E/D(S)}$ & Time complexity for the execution of the symmetric encryption/decryption, $3.85 ms$ \\
        \hline
        $T_h$ & Time complexity for the execution of the hash function, $0.0046 ms$ \\
        \hline
        $T_{ECPA}$ & Time  complexity for the execution of the elliptic curve (ECC) point addition, $0.0288 ms$ \\
        \hline
        $T_{EMod}$ & Time complexity for the execution of the modular exponential operation, $3.85 ms$ \\
        \hline
        \end{tabular}
    \end{center}
    \label{table:crypto-time}
\end{table*}
\begin{table*}[htbp]
\caption{Comparison of computational overhead}
    \begin{center}
    \begin{tabular}{|c|c||c|}
        \hline
        \textbf{Protocol}&\textbf{Overhead}&\textbf{Estimated Time} \\
        \hline
        LAKEE & $6T_{ECPM} + 3T_{E/D(s)} + 34T_h$ & 25.06ms \\
        \hline
        ECC-CoAP \cite{Majumder} & $6T_{ECPM} + 4T_{E/D(s)} + 5T_h$ & 28.779ms \\
        \hline
        Dey and Hossain scheme \cite{dey2019session} & $4T_{EMod} + 6T_{E/D(s)} + 6T_h$ & 38.527ms \\
        \hline
        \end{tabular}
    \end{center}
    \label{table:comput}
\end{table*}
\par
$\textit{overhead = } 6T_{ECPM} + 3T_{E/D(s)} + 2T_h + 32T_h= 13.356 + 11.55 + 0.0092 + 0.1472 = 25.06 \,ms$
\par
Table \ref{table:comput} shows the efficiency of LAKEE in comparison with other protocols seeking the same goal. 
\subsection{Communication Overhead}
\begin{table*}[htbp]
\caption{Comparison of communication overhead}
    \begin{center}
    \begin{tabular}{|c|c|}
        \hline
        \textbf{Protocol}&\textbf{Number of Messages} \\
        \hline
        LAKEE & 3\\
        \hline
        ECC-CoAP \cite{Majumder} & 4 \\
        \hline
        LESS \cite{less} & 4 \\
        \hline
        Dey and Hossain scheme \cite{dey2019session} & 5\\
        \hline
        DTLS Handshake \cite{RFC6347} & 6\\
        \hline
        \end{tabular}
    \end{center}
    \label{table:commu}
\end{table*}
\par For calculating the communication overhead in bits, we have to specify the length of each element. Ideally, the client ID is 64-bit long, and timestamps are 32 bits long. If we use the curve \textit{Ed448}, $server\_rand$ and $client\_rand$ are 224 bits long. If we use other curves with a 128-bit secury level, $server\_rand$ and $client\_rand$ would normally be 128 bits long. The MAC, which the ciphersuite appends to the plaintext, has a fixed size of 128 bits. We use SHA-1 as a hash function for the client's ID with a 160-bit long output. The ciphertext would be the same size as the plaintext. The message sizes are calculated below.
\begin{itemize}
    \item first message: $64+160+224+32+128 = 608 \, bits$
    \item second message:
    $224+224+32+128 = 608 \, bits$
    \item third message:
    $224+32+128 = 384 \, bits$
    \item total:
    $1600 \, bits$
\end{itemize}
\par In practice, each message should include a header not necessarily specified in the protocol. Therefore, besides the size of the messages in bits, the number of messages corresponds to the protocol's efficiency as well. LAKEE requires 3 messages for session key generation and authentication, which is less than its alternatives. Table \ref{table:commu} compares the total number of messages required by LAKEE and its alternatives. 
\section{Conclusion}
In this paper, we proposed an efficient and lightweight authenticated key exchange protocol for the setting of IoT by employing elliptic curve cryptography. We addressed the performance and security shortcomings of the existing protocols. The results are promising.  and our protocol, LAKEE, is proven to be more efficient than alternative protocols. LAKEE is implemented as an application protocol using CoAP. The security of it is analyzed formally and using the AVISPA security tool. 
\section{Future Work}
One contribution to this work would be implementing the protocol in the transport layer and enabling DTLS to use it for session key generation instead of DTLS handshake.
Another contribution would be securing LAKEE against distributed denial of service attacks (DDoS). In the case of a DDoS attack, blocking a certain IP range would not restrict adversaries' access to the network. 

\bibliographystyle{unsrt}  
\bibliography{lakee}

\begin{thebibliography}{10}

\bibitem{MQTTIssues}
Muneer O.~Bani Yassein, Mohammed~Q. Shatnawi, Shadi Aljwarneh, and Razan
  Al-Hatmi.
\newblock Internet of things: Survey and open issues of mqtt protocol.
\newblock {\em 2017 International Conference on Engineering \& MIS (ICEMIS)},
  pages 1--6, 2017.

\bibitem{restfulCoAP}
Hoai~Viet Nguyen and Luigi~Lo Iacono.
\newblock Rest-ful coap message authentication.
\newblock {\em 2015 International Workshop on Secure Internet of Things
  (SIoT)}, pages 35--43, 2015.

\bibitem{standardIoT}
Maria~Rita Palattella, Nicola Accettura, Xavier Vilajosana, Thomas Watteyne,
  Luigi~Alfredo Grieco, Gennaro Boggia, and Mischa Dohler.
\newblock Standardized protocol stack for the internet of (important) things.
\newblock {\em IEEE Communications Surveys Tutorials}, 15(3):1389--1406, 2013.

\bibitem{steiner1988kerberos}
Jennifer~G Steiner, B~Clifford Neuman, and Jeffrey~I Schiller.
\newblock Kerberos: An authentication service for open network systems.
\newblock In {\em Usenix Winter}, pages 191--202. Citeseer, 1988.

\bibitem{fischlin2014multi}
Marc Fischlin and Felix G{\"u}nther.
\newblock Multi-stage key exchange and the case of google's quic protocol.
\newblock In {\em Proceedings of the 2014 ACM SIGSAC Conference on Computer and
  Communications Security}, pages 1193--1204, 2014.

\bibitem{SAKE}
Gildas Avoine, S{\'e}bastien Canard, and Lo{\"i}c Ferreira.
\newblock Symmetric-key authenticated key exchange (sake) with perfect forward
  secrecy.
\newblock In Stanislaw Jarecki, editor, {\em Topics in Cryptology -- CT-RSA
  2020}, pages 199--224, Cham, 2020. Springer International Publishing.

\bibitem{HTTP-COAP}
M.~Harish, R.~Karthick, R.~Mohan~Rajan, and V.~Vetriselvi.
\newblock Securing coap through payload encryption: Using elliptic curve
  cryptography.
\newblock In Amit Kumar and Stefan Mozar, editors, {\em ICCCE 2018}, pages
  497--511, Singapore, 2019. Springer Singapore.

\bibitem{SOAP}
Guido Moritz, Frank Golatowski, and Dirk Timmermann.
\newblock A lightweight soap over coap transport binding for resource
  constraint networks.
\newblock In {\em 2011 IEEE Eighth International Conference on Mobile Ad-Hoc
  and Sensor Systems}, pages 861--866, 2011.

\bibitem{aiocoAP}
Christian Amsüss.
\newblock aiocoap, 2014.

\bibitem{Pan}
Menghan Pan, Daojing He, Xuru Li, Sammy Chan, Emmanouil Panaousis, and Yun Gao.
\newblock A lightweight certificateless non-interactive authentication and key
  exchange protocol for iot environments.
\newblock In {\em 2021 IEEE Symposium on Computers and Communications (ISCC)},
  pages 1--7, 2021.

\bibitem{Kalra}
Sheetal Kalra and Sandeep~K. Sood.
\newblock Secure authentication scheme for iot and cloud servers.
\newblock {\em Pervasive and Mobile Computing}, 24:210--223, 2015.
\newblock Special Issue on Secure Ubiquitous Computing.

\bibitem{Kumari}
Saru Kumari, Marimuthu Karuppiah, Ashok~Kumar Das, Xiong Li, Fan Wu, and Neeraj
  Kumar.
\newblock A secure authentication scheme based on elliptic curve cryptography
  for {IoT} and cloud servers.
\newblock {\em The Journal of Supercomputing}, 74(12):6428--6453, December
  2018.

\bibitem{adesh}
Adesh Kumari, Vinod Kumar, M.~YahyaAbbasi, and Mansaf Alam.
\newblock The cryptanalysis of a secure authentication scheme based on elliptic
  curve cryptography for iot and cloud servers.
\newblock In {\em 2018 International Conference on Advances in Computing,
  Communication Control and Networking (ICACCCN)}, pages 321--325, 2018.

\bibitem{shin2016MQTT}
SeongHan Shin, Kazukuni Kobara, Chia-Chuan Chuang, and Weicheng Huang.
\newblock A security framework for mqtt.
\newblock In {\em 2016 IEEE Conference on Communications and Network Security
  (CNS)}, pages 432--436. IEEE, 2016.

\bibitem{shin2012augpake}
S~Shin and Kazukuni Kobara.
\newblock Efficient augmented password-only authentication and key exchange for
  ikev2.
\newblock Technical report, 2012.

\bibitem{YehWirelessECC}
Hsiu-Lien Yeh, Tien-Ho Chen, Pin-Chuan Liu, Tai-Hoo Kim, and Hsin-Wen Wei.
\newblock A secured authentication protocol for wireless sensor networks using
  elliptic curves cryptography.
\newblock {\em Sensors}, 11(5):4767--4779, 2011.

\bibitem{dey2019session}
Shreya Dey and Ashraf Hossain.
\newblock Session-key establishment and authentication in a smart home network
  using public key cryptography.
\newblock {\em IEEE Sensors Letters}, 3(4):1--4, 2019.

\bibitem{less}
Abhijan Bhattacharyya, Tulika Bose, Soma Bandyopadhyay, Arijit Ukil, and Arpan
  Pal.
\newblock Less: Lightweight establishment of secure session: A cross-layer
  approach using coap and dtls-psk channel encryption.
\newblock In {\em 2015 IEEE 29th International Conference on Advanced
  Information Networking and Applications Workshops}, pages 682--687, 2015.

\bibitem{Majumder}
Suman Majumder, Sangram Ray, Dipanwita Sadhukhan, Muhammad~Khurram Khan, and
  Mou Dasgupta.
\newblock Ecc-coap: Elliptic curve cryptography based constraint application
  protocol for internet of things.
\newblock {\em Wireless Personal Communications}, 116(3):1867--1896, Feb 2021.

\bibitem{alhaj2018constrained}
A~Alhaj.
\newblock Constrained application protocol (coap) for the iot.
\newblock In {\em IOT SEMINAR, HIGH INTEGRITY SYSTEM, FRANKFURT UNIVERSITY OF
  APPLIED SCIENCE, Frankfurt}, 2018.

\bibitem{ECCMiller}
Victor~S. Miller.
\newblock Use of elliptic curves in cryptography.
\newblock In {\em Advances in Cryptology}, CRYPTO '85, page 417–426, Berlin,
  Heidelberg, 1985. Springer-Verlag.

\bibitem{ECCKoblitz}
N.~Koblitz.
\newblock {\em Towards a Quarter-Century of Public Key Cryptography}.
\newblock Designs, codes and cryptography. Springer US, 2000.

\bibitem{RFC4279}
P.~Eronen and H.~Tschofenig.
\newblock Pre-shared key ciphersuites for transport layer security (tls).
\newblock RFC 4279, RFC Editor, December 2005.
\newblock \url{http://www.rfc-editor.org/rfc/rfc4279.txt}.

\bibitem{RFC6347}
E.~Rescorla and N.~Modadugu.
\newblock Datagram transport layer security version 1.2.
\newblock RFC 6347, RFC Editor, January 2012.
\newblock \url{http://www.rfc-editor.org/rfc/rfc6347.txt}.

\bibitem{hamburg2015ed448}
Mike Hamburg.
\newblock Ed448-goldilocks, a new elliptic curve.
\newblock {\em Cryptology ePrint Archive}, 2015.

\bibitem{RFC7748}
A.~Langley et~al.
\newblock Elliptic curves for security.
\newblock RFC 7748, RFC Editor, January 2016.
\newblock \url{http://www.rfc-editor.org/rfc/rfc7748.txt}.

\bibitem{SafeCurves}
Daniel~J. Bernstein and Tanja Lange.
\newblock Safecurves: choosing safe curves for elliptic-curve cryptography.

\bibitem{PBKDF2RFC2898}
B.~Kaliski.
\newblock Password-based cryptography specification version 2.0.
\newblock RFC 2898, RFC Editor, September 2000.
\newblock \url{http://www.rfc-editor.org/rfc/rfc2898.txt}.

\bibitem{dolev-yao}
Danny Dolev and Andrew Yao.
\newblock On the security of public key protocols.
\newblock {\em IEEE Transactions on information theory}, 29(2):198--208, 1983.

\bibitem{team2006avispa}
TA~Team et~al.
\newblock Avispa v1. 1 user manual.
\newblock {\em Information society technologies programme (June 2006)
  http://avispa-project. org}, 2006.

\bibitem{heen2008industrialspan}
Olivier Heen, Thomas Genet, St{\'e}phane Geller, and Nicolas Prigent.
\newblock An industrial and academic joint experiment on automated verification
  of a security protocol.
\newblock In {\em Mobile and Wireless Networks Security}, pages 39--53. World
  Scientific, 2008.

\bibitem{Mahmood}
Khalid Mahmood, Shehzad~Ashraf Chaudhry, Husnain Naqvi, Saru Kumari, Xiong Li,
  and Arun~Kumar Sangaiah.
\newblock An elliptic curve cryptography based lightweight authentication
  scheme for smart grid communication.
\newblock {\em Future Gener. Comput. Syst.}, 81(C):557–565, apr 2018.

\bibitem{Burrows}
Michael Burrows, Martin Abadi, and Roger Needham.
\newblock A logic of authentication.
\newblock {\em ACM Trans. Comput. Syst.}, 8(1):18–36, feb 1990.

\bibitem{syverson1996unified}
Paul~F Syverson and Paul~C Van~Oorschot.
\newblock A unified cryptographic protocol logic.
\newblock Technical report, NAVAL RESEARCH LAB WASHINGTON DC, 1996.

\bibitem{abadi1991semantics}
Martin Abadi and Mark~R Tuttle.
\newblock A semantics for a logic of authentication.
\newblock In {\em Proceedings of the tenth annual ACM symposium on Principles
  of distributed computing}, pages 201--216, 1991.

\bibitem{gong1990reasoning}
Li~Gong, Roger~M Needham, and Raphael Yahalom.
\newblock Reasoning about belief in cryptographic protocols.
\newblock In {\em IEEE Symposium on Security and Privacy}, volume 1990, pages
  234--248. Citeseer, 1990.

\bibitem{oorschotVO}
Paul van Oorschot.
\newblock Extending cryptographic logics of belief to key agreement protocols.
\newblock In {\em Proceedings of the 1st ACM Conference on Computer and
  Communications Security}, CCS '93, page 232–243, New York, NY, USA, 1993.
  Association for Computing Machinery.

\bibitem{TheLogicofAuthenProt}
Paul Syverson and Iliano Cervesato.
\newblock The logic of authentication protocols.
\newblock In Riccardo Focardi and Roberto Gorrieri, editors, {\em Foundations
  of Security Analysis and Design}, pages 63--137, Berlin, Heidelberg, 2001.
  Springer Berlin Heidelberg.

\bibitem{Kilinc}
H.~Hakan Kilinc and Tugrul Yanik.
\newblock A survey of sip authentication and key agreement schemes.
\newblock {\em IEEE Communications Surveys Tutorials}, 16(2):1005--1023, 2014.

\end{thebibliography}

\end{document}